\documentstyle[aps,prl,twocolumn,psfig]{revtex}
\begin{document}
\preprint{\vbox{\hbox{ISU-HEP-99-7}}, \vbox{\hbox{UCB-PTH-99/42}},
  \vbox{LBNL-44308}}
\draft
\title{CP Violation in Hyperon Decays from Supersymmetry}
\author{Xiao-Gang He$^1$, Hitoshi~Murayama$^{2,3}$, 
Sandip Pakvasa$^4$ and G. Valencia$^5$
}

\address{
$^1$Department of Physics, National Taiwan University, Taipei, 10674
\\
$^2$Department of Physics, University of California, Berkeley, CA 94720
\\
$^3$Theoretical Physics Group, Lawrence Berkeley National Laboratory, 
Berkeley, CA 94720
\\
$^4$Department of Physics and Astronomy, University of Hawaii at Manoa,
Honolulu,
HI 96822\\
$^5$Department of Physics, Iowa State University, Ames, Iowa 50011
}

\date{\today}
\maketitle
\begin{abstract}

It was pointed out recently that supersymmetry can generate 
flavor-changing gluonic dipole operators with sufficiently large 
coefficients to dominate the observed value of 
$\epsilon^\prime/\epsilon$.  We point out that the 
same operators contribute to direct CP violation in hyperon decay 
and can generate a CP violating asymmetry $A(\Lambda^0_-)$ 
in the range probed by the current E871 experiment. 
Interestingly, models that naturally reproduce the relation 
$\lambda = \sqrt{m_{d}/m_{s}}$ do not generate 
$\epsilon^{\prime}/\epsilon$ but could lead to an $A(\Lambda^{0}_-)$ 
of $O(10^{-3})$. 

\end{abstract}

\narrowtext

The origin of CP violation remains one of the outstanding problems 
in particle physics. Until recently the only observation of  CP violation 
was in the neutral kaon mixing, with a value of 
$\epsilon \approx 2.27\times 10^{-3} \exp(i\pi/4)$ \cite{1,1a}. The KTeV and
NA48 collaborations have now reported observations of direct  
CP violation in the neutral kaon decay amplitudes \cite{2}, with 
the world average value being 
${\rm Re}(\epsilon'/\epsilon) = (21.2 \pm 4.6)\times 10^{-4}$ \cite{WA}. 

Although this result is not inconsistent with the standard model 
prediction, it can be used to constrain other  
models of CP violation \cite{susy,neuka,other}. In particular, 
it has been found that there can be large supersymmetric 
contributions to $\epsilon'/\epsilon $ \cite{susy,neuka}. 
Depending on which new contributions are large, there 
are different consequences for other processes such as 
rare kaon decays \cite{buras} and hyperon decays.

In this letter we concentrate on the supersymmetric 
scenario in which the gluonic dipole operators can 
have large coefficients. In this case there are potentially 
large contributions to both $\epsilon'/\epsilon$ \cite{susy} 
and to the CP violating asymmetry $A(\Lambda^0_-)$ 
in hyperon non-leptonic decays.

Experiment E871 at Fermilab is expected to reach a sensitivity 
of $2\times 10^{-4}$ for the observable 
($A(\Lambda^0_-) + A(\Xi^-_-)$) \cite{expe}.  
The CP violating asymmetry $A(\Lambda^0_-)$ compares the decay 
parameter $\alpha$ from the reaction $\Lambda^0 \rightarrow p \pi^-$ 
to the corresponding parameter $\bar\alpha$ in 
$\bar\Lambda^0 \rightarrow \bar p \pi^+$ whereas $A(\Xi^-_-)$ is 
the corresponding asymmetry for the mode $\Xi^- \rightarrow 
\Lambda^0 \pi^-$. These asymmetries have a very simple form 
when one neglects the small $\Delta I =3/2$ amplitude, for 
example \cite{3},
\begin{eqnarray}
A(\Lambda^0_-) &=& {\alpha+\bar \alpha\over \alpha -\bar \alpha}
\approx -\tan(\delta_{11} - \delta_1) \sin (\phi_p -\phi_s),
\end{eqnarray}
where $\delta_{1} = 6^\circ$, $\delta_{11}=-1.1^\circ$  
are the final state $\pi N$ interaction phases for S and P wave amplitudes 
with $I =1/2$, respectively \cite{roper}. $\phi_{s,p}$ are the
corresponding CP violating weak phases. Recent
calculations suggest that the strong scattering phases in 
the $\Lambda^0 \pi$ final state of the $\Xi$ decay are small \cite{stph}, 
and, therefore, the current theoretical prejudice is that 
$|A(\Lambda^0_-)|$ will dominate the measurement. The standard model 
prediction for this quantity is around $3\times 10^{-5}$, albeit with large 
uncertainty \cite{3,4}. This suggests that a non-zero measurement 
by E871 will be an indication for new physics. 

A model independent study of new CP violating interactions 
has shown that $A(\Lambda^0_-)$ 
could be ten times larger than in the standard model and within 
reach of E871 \cite{hvop}. A particular example of an operator in which 
$A(\Lambda^0_-)$ can be this large is precisely the gluonic 
dipole operator \cite{hvop}. The results of E871, 
therefore, can have a direct impact on supersymmetric models. 

The short distance effective Hamiltonian for the gluonic dipole
operator of interest is,
\begin{eqnarray} 
{\cal H}_{\it eff} &=& C_{g} 
{g_s\over 16\pi^2} m_s \bar d \sigma_{\mu\nu} G^{\mu\nu}_a t^a
(1+\gamma_5) s \nonumber \\
& &+ 
\tilde{C}_{g} 
{g_s\over 16\pi^2} m_s \bar d \sigma_{\mu\nu} G^{\mu\nu}_a t^a
(1-\gamma_5) s ~+~{\rm h.c.},
\label{effh}
\end{eqnarray}
where $Tr(t^at^b) = \delta^{ab}/2$, and the Wilson coefficients 
$C_{g}$ and $\tilde{C}_{g}$ that occur in supersymmetry 
can be found in the literature \cite{ggms}, they are
\begin{equation}
C_{g} = (\delta^d_{12})_{LR} 
{\alpha_s \pi \over  m_{\tilde g} m_s} 
G_0(x) \;,\;
\tilde{C}_{g} = (\delta^d_{12})_{RL} 
{\alpha_s \pi \over  m_{\tilde g} m_s} 
G_0(x) .
\label{wilco}
\end{equation}
The parameters $\delta_{12}^{d}$ characterize the mixing in 
the mass insertion approximation \cite{ggms}, and 
$x=m_{\tilde g}^2/m_{\tilde q}^2$, 
with $m_{\tilde g}$, $m_{\tilde q}$ being the gluino and average  
squark masses, respectively. The loop function is given by, 
\begin{equation}
        G_0(x)=x 
        {22-20x-2x^2+(16x-x^2+9)\log x
        \over 3 (x-1)^4}.
\end{equation}
Ref.~\cite{buras} has noted that, in this form, $G_0(1)=-5/18$ 
and the function does not depend strongly on $x$. 
The effect of QCD corrections is to multiply the Wilson 
coefficients by \cite{qcdcorr}
\begin{equation}
\eta=\biggl({\alpha_s(m_{\tilde{g}})\over \alpha_s(m_t)}
\biggr)^{2\over 21}
\biggl({\alpha_s(m_t)\over \alpha_s(m_b)}
\biggr)^{2\over 23}
\biggl({\alpha_s(m_b)\over \alpha_s(m_c)}
\biggr)^{2\over 25} .
\end{equation}

To calculate the weak phases we adopt the usual procedure 
of taking the real part of the amplitudes from experiment 
and of using a model for the hadronic matrix elements 
to obtain the imaginary part. We use the MIT bag model matrix elements 
of Ref.~\cite{3,ponce} to find for the weak phases
\begin{eqnarray}
\phi_s &= & -2.9 \times 10^7~{\rm GeV} \nonumber\\
& &{\alpha_s \over 32 \pi}{\eta\over m_{\tilde{g}}}G_0(x)
{\rm Im}\biggl((\delta_{12}^d)_{LR}-(\delta_{12}^d)_{RL}\biggr)
B_s,
\label{bagphases}\\
\phi_p &= & -3.4 \times 10^7~{\rm GeV} \nonumber\\
& &{\alpha_s \over 32 \pi}{\eta\over m_{\tilde{g}}}G_0(x)
{\rm 
Im}\biggl((\delta_{12}^d)_{LR}+(\delta_{12}^d)_{RL}\biggr)B_p.
\label{bagphasep}
\end{eqnarray}
We have introduced the parameters $B_s$ and $B_p$ to quantify 
the uncertainty in these matrix elements. We then find, 
\begin{eqnarray}
        \lefteqn{
        A(\Lambda^0_-)_{SUSY} =
        \left({\alpha_s(m_{\tilde{g}}) \over \alpha_s(500~{\rm GeV})}
        \right)^{23\over 21}
        \left({500~{\rm GeV}\over m_{\tilde{g}}}\right) 
        {G_0(x)\over G_0(1)} 
        } \nonumber \\
        &&
        \left((2.0 B_p - 1.7 B_s) {\rm Im}(\delta_{12}^d)_{LR}
        +(2.0 B_p + 1.7 B_s) {\rm Im}(\delta_{12}^d)_{RL} \right).
        \nonumber \\
\label{cpasym}
\end{eqnarray}
The matrix element of the gluonic dipole operator of 
Eq.~(\ref{effh}) between two baryon states is calculated 
with the MIT bag model in Ref.~\cite{ponce}, 
and we assume that this result is accurate to within factors of two. 
The S-wave hyperon decay amplitude is then obtained 
by using a soft pion theorem which can have $20-30\%$ 
corrections. The P-wave hyperon decay amplitude is obtained by 
considering baryon and kaon pole diagrams. 
A leading order calculation of the dominant, CP conserving, 
P-wave amplitudes in terms of (octet) baryon poles alone works 
reasonably well for $\Lambda^0$ decays. However, additional contributions 
are needed to explain the P-waves in other hyperon decays \cite{donbook}, 
and the first non-leading corrections to the $\Lambda^0$ decay amplitude 
are large. An example of an additional contribution is the kaon pole, 
which in Eq.~(\ref{bagphasep}) accounts for about 20\% 
of the P-wave phase.  To reflect these uncertainties in 
our numerical analysis we use $0.5 < B_s < 2.0$, while allowing 
$B_p$ to vary in the range $0.7 B_s < B_p < 1.3 B_s$.  

In a general supersymmetric model there are also contributions 
to the imaginary parts of the Wilson coefficients of four-quark 
operators. Of these, the dominant contribution to the CP asymmetry 
in hyperon decays (within the standard model) 
is due to $O_6$ \cite{4}. We have checked numerically, that 
SUSY contributions to $C_6$ (as well as to $C_{3,4,5,7}$) are much 
smaller than those in Eq.~(\ref{cpasym}), for a parameter range 
similar to that considered in Ref.~\cite{ggms}.

Although the asymmetry $A(\Lambda^0_-)$ is due to the same 
$|\Delta S| = 1$ interaction responsible for $\epsilon'/\epsilon$, 
the two observables are qualitatively different. 
For $\epsilon'/\epsilon$, both the $\Delta I =1/2$ and the $\Delta I=3/2$ 
amplitudes are equally important, whereas for $A(\Lambda^0_-)$ 
only the $\Delta I =1/2$ amplitude is important. In this case, the 
interference necessary for CP violation takes place between 
$S$ and $P$ waves within the $\Delta I =1/2$ transition. 
This sensitivity to differences between $S$ and $P$ waves accounts 
for the different coefficients multiplying 
$(\delta_{12}^d)_{LR}$ and $(\delta_{12}^d)_{RL}$ respectively in 
Eq.~(\ref{cpasym}). For this same reason, supersymmetric scenarios 
in which $\epsilon^\prime$ is enhanced through $\Delta I =3/2$ operators 
\cite{neuka,buras} do not enhance $A(\Lambda^0_-)$.

In order to quantify $A(\Lambda^0_-)$ in supersymmetric models where 
the operators in Eq.~(\ref{effh}) have large coefficients, we 
compare Eq.~(\ref{cpasym}) with their contributions
to $\epsilon^\prime/\epsilon$ \cite{buras}, 
\begin{eqnarray}
        \lefteqn{
\biggl({\epsilon^\prime \over \epsilon}\biggr)_{SUSY} =
\biggl({\alpha_s(m_{\tilde{g}}) \over \alpha_s(500~{\rm GeV})}
\biggr)^{23\over 21}
\biggl({500~{\rm GeV}\over m_{\tilde{g}}}\biggr) 
{G_0(x)\over G_0(1)} B_G 
} \nonumber \\
& &\biggl( {158~{\rm MeV}\over m_s(m_c)+m_d(m_c)} \biggr)
58\  
{\rm Im}\biggl( (\delta_{12}^d)_{LR}-(\delta_{12}^d)_{RL} \biggr)\ .
\label{epspsusy}
\end{eqnarray}
To obtain this expression, 
Ref.~\cite{buras} uses the $K\rightarrow \pi\pi$ matrix element from 
a chiral quark model calculation in Ref.~\cite{14} and uses the parameter  
$B_G$ to quantify the hadronic uncertainty. We use the range 
$0.5 < B_G < 2$ motivated by the bag model result of Ref.~\cite{donhol} 
and the dimensional analysis estimate of Ref.~\cite{nhv}. 
It is interesting to 
note that $(\epsilon'/\epsilon)_{SUSY}$ depends on the same combination of the 
mass insertion parameters as the weak phase $\phi_{s}$ in 
Eq.~(\ref{bagphases}).  We require $(\epsilon'/\epsilon)_{SUSY}$ to be 
equal to the observed value ({\it i.e.}\/, $\epsilon'/\epsilon$ dominated by 
supersymmetry) or less ({\it i.e.}\/, $\epsilon'/\epsilon$ dominated by the  
standard model). 

Comparing Eqs.~(\ref{cpasym})~and~(\ref{epspsusy}) one sees that 
$\epsilon^\prime/\epsilon$ and $A(\Lambda^0_-)$ 
are proportional to different combinations 
of the coefficients $(\delta_{12}^d)_{LR}$ and $(\delta_{12}^d)_{RL}$. 
For this reason one cannot determine the allowed range for  
$A(\Lambda^0_-)$ solely in terms 
of $\epsilon^\prime/\epsilon$.  In what follows, we consider the three cases:
a) ${\rm Im}(\delta_{12}^d)_{RL}=0$,
b) ${\rm Im}(\delta_{12}^{d})_{LR}=0$, and 
c) ${\rm Im}(\delta_{12}^d)_{RL}= 
{\rm Im}(\delta_{12}^d)_{LR}$ motivated below.

It is useful to recall the origin of the mass insertion parameters 
$(\delta_{12}^d)_{RL}$ and $(\delta_{12}^d)_{LR}$.  They are the 
mismatch between the quark mass matrix and left-right mass-squared 
matrix for down-type squarks (we restrict our discussion to the first 
and second generations).  In many theories of flavor with 
approximate flavor symmetries, the Cabibbo angle originates in the 
down-quark sector, and we find the mass matrix to be of the form
\begin{equation}
        M_{d} = \left( \begin{array}{cc}
                a m_{s}\lambda^{2} & m_{s} \lambda \\
                b m_{s} \lambda & m_{s}
        \end{array}
        \right) ,
        \label{eq:Md}
\end{equation}
where $a$ and $b$ are $O(1)$ coefficients and $\lambda$ is the sine 
of the Cabibbo 
angle.  The (2,2) element is nothing but the strange quark mass itself 
(ignoring $O(\lambda^{2})$ corrections), and the (1,2) element is 
fixed by the requirement that the Cabibbo angle is reproduced.  The 
down quark mass is given by $m_{d} = (a-b) \lambda^{2}$.  A case of 
$a=0$ and $b=-1$ naturally reproduces the phenomenologically 
successful relation $\lambda \simeq \sqrt{m_{d}/m_{s}}$ and deserves a 
special attention.  This arises if the off-diagonal elements originate 
in an anti-symmetric matrix such as in U(2) model \cite{U(2)}.  The 
$\lambda$ dependence of each element is a consequence of the 
approximate flavor symmetry, but the constants $a$ and $b$
cannot be determined by symmetry considerations alone and hence are
model-dependent.  The same approximate flavor symmetry constrains the 
form of the left-right mass-squared matrix.  Therefore, the left-right 
mass-squared matrix for down-type squarks is
\begin{equation}
        M^{2d}_{LR} = m_{SUSY} \left( \begin{array}{cc}
                \tilde{a} m_{s}\lambda^{2} & \tilde{c}m_{s} \lambda \\
                \tilde{b} m_{s} \lambda & \tilde{d} m_{s}
        \end{array}
        \right),
        \label{eq:LR}
\end{equation}
where $m_{SUSY}$ is the typical supersymmetry breaking scale which we 
take to be the same as the down-type squark mass, and $\tilde{a}$, 
$\tilde{b}$, $\tilde{c}$, and $\tilde{d}$ are $O(1)$ numbers and can 
be complex.  The U(2) model gives $\tilde{b} = - \tilde{c}$.

After diagonalizing the quark mass matrix Eq.~(\ref{eq:Md}), the 
left-right mass-squared matrix becomes
\begin{equation}
        m_{SUSY} m_{s} \left( \begin{array}{cc}
                (\tilde{a}-b\tilde{c}-\tilde{b}+b\tilde{d}) \lambda^{2} &
                (\tilde{c} - \tilde{d}) \lambda\\
                (\tilde{b} - b \tilde{d}) \lambda & \tilde{d}
        \end{array}
        \right).
        \label{eq:LR2}
\end{equation}
Unless special relations hold between $O(1)$ coefficients, there 
remain off-diagonal elements which contribute to flavor-changing 
neutral currents.  The mass insertion parameters for $s\rightarrow d$ 
transition are defined as
\begin{eqnarray}
        & & (\delta_{12}^{d})_{LR} = \frac{m_{s} (\tilde{c} - 
        \tilde{d})}{m_{SUSY}}, \nonumber \\
        & &(\delta_{12}^{d})_{RL} = (\delta_{21}^{d})_{LR}^{*} =
        \frac{m_{s} (\tilde{b} - b \tilde{d})^{*}}{m_{SUSY}} .
        \label{eq:delta}
\end{eqnarray}
It is amusing that the size of the mass insertion parameters given here 
generates $\epsilon'$ according to Eq.~(\ref{epspsusy}) at the 
observed level for $m_{SUSY} \sim 500$~GeV and a phase of $O(1)$.  

The case a), of ${\rm Im}(\delta_{12}^{d})_{LR} \neq 0$ and 
${\rm Im}(\delta_{12}^{d})_{RL}=0$, 
corresponds to the choice $a=1$, $b=0$ in the quark mass matrix
Eq.~(\ref{eq:Md}) and its counter part in the squark mass matrix
Eq.~(\ref{eq:LR}) $\tilde{b}=0$ is also likely to be zero in this case.  
We still
expect $\tilde{c}$, $\tilde{d}$ to be $O(1)$ and this case is the most
conservative one.  The case b) is the other possible limit 
where $\tilde{c}-\tilde{d}$ happens to have a negligible imaginary 
part.  ${\rm Im}(\tilde{b}-b\tilde{d})$ can still generate an 
interesting contribution to $\epsilon'$, while $A(\Lambda^{0}_{-})$ can 
be much larger in this case.  Finally, the case c) 
${\rm Im}(\delta_{12}^d)_{RL}={\rm Im}
(\delta_{12}^d)_{LR}$ is motivated by the phenomenological relation 
$\lambda \simeq \sqrt{m_{d}/m_{s}}$ and hence $a=0$, $b=-1$.  The 
anti-symmetry in $M_{d}$ could imply the anti-symmetry in 
$M^{2d}_{LR}$, and hence $\tilde{b} = - \tilde{c}$.  This is indeed 
what happens in the U(2) model of flavor \cite{U(2)}.  In this case, 
${\rm Im}(\delta_{12}^{d})_{LR} = m_{s} {\rm Im} 
(-\tilde{b}-\tilde{d})/m_{SUSY}$, while ${\rm Im} (\delta_{12}^{d})_{RL} 
= m_{s }{\rm Im} (\tilde{b} + \tilde{d})^{*}/m_{SUSY} = m_{s}{\rm Im} 
(-\tilde{b}-\tilde{d})/m_{SUSY} = {\rm Im}(\delta_{12}^{d})_{LR}$.  
Therefore, there is no parity violation in the CP-violating part of the 
operators and hence the contribution to $\epsilon'$ identically 
vanishes \cite{footnote}.  
In this case, the only constraint on the size of 
$A(\Lambda^{0}_-)$ comes from $\epsilon$ as we will discuss below.

The operators in Eq.~(\ref{effh}) also contribute to $\epsilon$ 
through long distance effects and we must check that this 
contribution is not too large. The simplest long distance  
contributions arise from $\pi^0$,~$\eta$~and~$\eta^\prime$ poles 
as noted in Ref.~\cite{9}. They yield,
\begin{eqnarray}
\bigl(\epsilon\bigr)_{SUSY} &=&
{1 \over \sqrt{2} m_K \Delta m}
{1\over m_K^2-m_\pi^2} 
\nonumber \\
& & {\rm Im}(\langle\pi^0|{\cal H}_{\it eff}|K^0\rangle)
\langle\pi^0|{\cal H}_{SM}|K^0\rangle \kappa .
\end{eqnarray}
In this expression $\Delta m$ is the $K_L-K_S$ mass 
difference and 
$\langle\pi^0|{\cal H}_{SM}|K^0\rangle \approx 2.6\times 10^{-8}~{\rm GeV}^2$ 
is
extracted from $K\rightarrow \pi\pi$ data. We get the matrix 
element $\langle\pi^0|{\cal H}_{\it eff}|K^0\rangle$ using the MIT bag model 
result \cite{ponce}. Finally, $\kappa$ quantifies the contributions  
of the different poles, $\kappa=1$ corresponding to the pion pole. 
In the model of Ref.~\cite{11} 
$\kappa \sim 0.2$ whereas the contribution of the $\eta^\prime$ 
alone gives $\kappa \sim -0.9$ \cite{3}. We use 
$0.2 < |\kappa| < 1.0$ and demand that this 
long distance contribution to $\epsilon$,  
\begin{eqnarray}
\bigl(\epsilon \bigr)_{SUSY}&=&
\biggl( {\alpha_s(m_{\tilde{g}}) \over \alpha_s(500~{\rm GeV})}
\biggr)^{23\over 21}
\biggl({500~{\rm GeV}\over m_{\tilde{g}} }\biggr)\ 
{\kappa\over 0.2}\ {G_0(x)\over G_0(1)}\nonumber \\
&& 6.4\  
{\rm Im}\biggl( (\delta_{12}^d)_{LR}+(\delta_{12}^d)_{RL} \biggr),
\label{epssusy}
\end{eqnarray}
be smaller than $2.3\times 10^{-3}$.  This leads to the constraint 
$|A(\Lambda^{0}_{-})| < 7.3 \times 10^{-4} B_{p}$.  Note that we allowed the range
$0.35<B_{p}<2.6$, and hence $|A(\Lambda^{0}_{-})|$ can be $O(10^{-3})$;
we cannot exclude it up to $1.9 \times 10^{-3}$. 
The constraint on the mass insertion parameters from the 
short-distance effect (e.g., box diagrams) is weaker: $({\rm Im}
(\delta_{12}^d)_{LR}^2)^{1/2} < 3.7 \times 10^{-3}$ for
$m_{\tilde{g}}=m_{\tilde{q}}=500$~GeV and
$(\delta_{12}^d)_{LR}=(\delta_{12}^d)_{RL}$ \cite{Ciuchini}.

\begin{figure}
  \centerline{\psfig{file=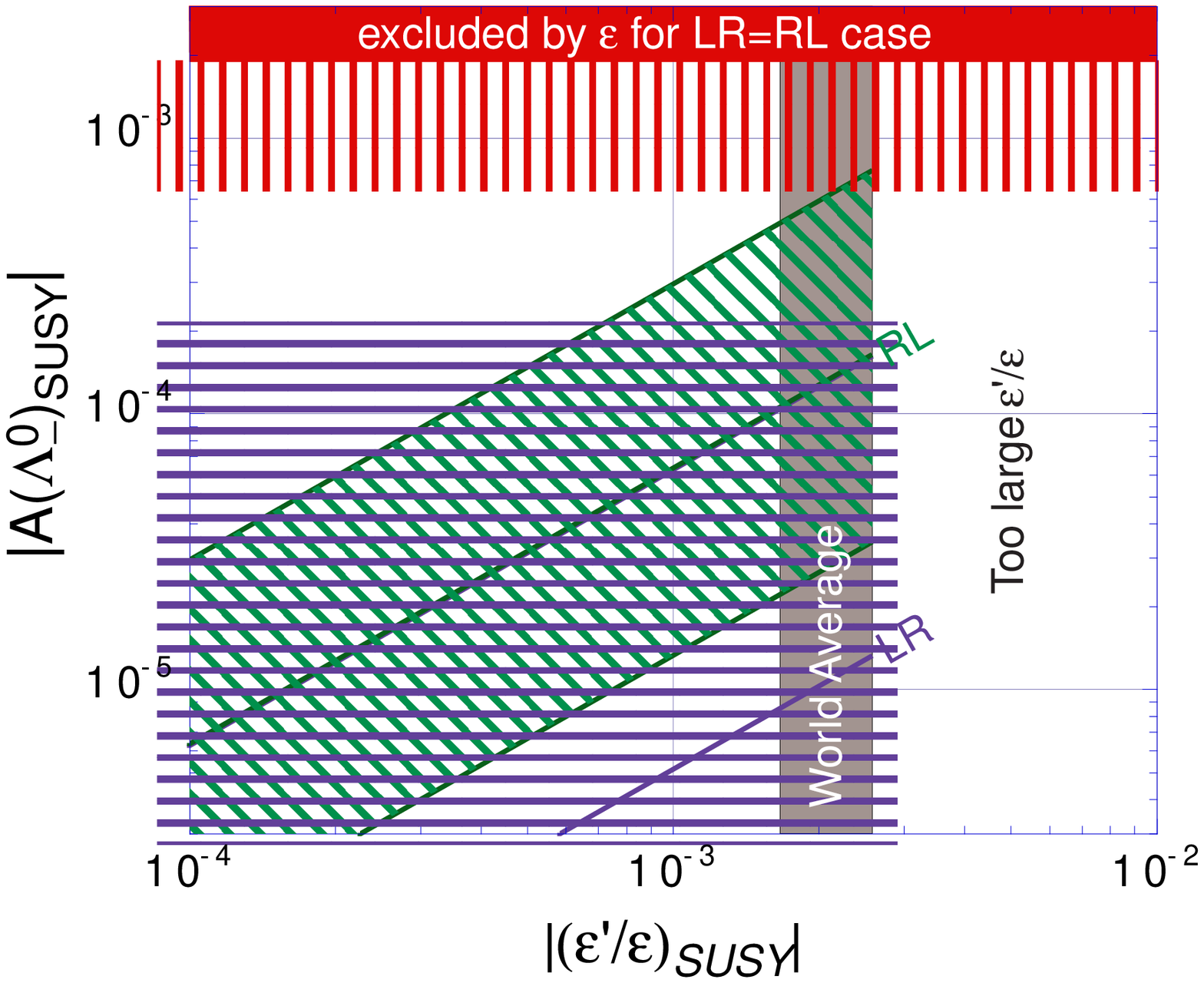,width=0.5\textwidth}}
  \caption{The allowed regions on $(|(\epsilon'/\epsilon)_{SUSY}|,$ 
    $|A(\Lambda^{0}_-)_{SUSY}|)$ parameter space for three cases: a)
    only ${\rm Im}(\delta^d_{12})_{LR}$ contribution, which is the
    conservative case (hatched horizontally), b) only ${\rm
      Im}(\delta^d_{12})_{RL}$ contribution (hatched diagonally), and
    c) ${\rm Im}(\delta^d_{12})_{LR}= {\rm Im}(\delta^d_{12})_{RL}$
    case which does not contribute to $\epsilon'$ and can give a large
    $|A(\Lambda^{0}_{-})|$ below the shaded region (or vertically
    hatched region for the central values of the matrix elements).  The
    last case is motivated by the relation $\lambda =
    \sqrt{m_{d}/m_{s}}$.  The vertical shaded band is the world
    average \protect\cite{WA} of $\epsilon'/\epsilon$.  The region to
    the right of the band is therefore not allowed.}
        \label{figure}
\end{figure}

The regions allowed by the three cases discussed above are shown in 
Fig.~\ref{figure}.  The case a) with LR contribution only is the 
horizontally-hatched region with the central value shown as a solid line, and 
the case b) with RL contribution only is the diagonally-hatched region 
with the central value shown as a solid line.  The shaded region at the top 
is excluded by the $\epsilon$ constraint, and is particularly important 
for case c) in 
which there is no contribution to $\epsilon'$.  It is 
interesting that the best motivated case c) allows a large 
asymmetry in hyperon decay.  The vertical band shows the world average 
for $\epsilon'/\epsilon$ and the region to the right of 
the band is, therefore, not allowed.

In summary, we have studied the supersymmetric contribution to 
CP violation in hyperon decays from gluonic dipole operators.  
We parameterize the hadronic uncertainties with the quantities 
$B_G$, $B_s$, $B_p$ and $\kappa$ which we allow to vary in 
reasonable ranges. We constrain the size of the coefficients of 
the gluonic dipole operators with the observed value of 
$\epsilon'$ and predict a range for $A(\Lambda^{0}_-)$ depending on 
whether the LR or the RL operator dominates.  We find  that the size of 
$A(\Lambda^{0}_-)$ can be within reach of the E871 experiment. 
Particularly interesting is the scenario c), which explains 
naturally the relation $\lambda = \sqrt{m_{d}/m_{s}}$. This scenario 
does not generate $\epsilon'$, but it can lead to an $A(\Lambda^{0}_-)$ 
as large as $10^{-3}$.

\acknowledgments

This work was supported in part by NSC of R.O.C. under grant number
NSC89-2112-M-002-016, by the Australian Research Council,
by DOE under contract numbers DE-AC03-76SF00098, DE-FG-03-94ER40833,
DEFG0292ER40730, and by NSF under grant PHY-95-14797.
G.V. thanks the theory group at SLAC for their 
hospitality while this work was completed.  We thank L.J. 
Hall and R. Barbieri for useful discussions.

\begin{thebibliography}{99}
\bibitem{WA} This is the average of E731, NA31, KTeV and NA48 with the
  error bar inflated to obtain $\chi^2/{\rm d.o.f.} = 1$ according to
  the Particle Data Group prescription.
\end{thebibliography}
\end{document}